\documentclass[prb,aps,twocolumn,showpacs,superscriptaddress]{revtex4-1}
\usepackage{amssymb}
\usepackage{amsmath}
\usepackage{mathptmx}
\usepackage[version=4]{mhchem}
\usepackage{hyperref}
\hypersetup{colorlinks=true, citecolor=blue, filecolor= black, linkcolor= blue, urlcolor= blue}
\usepackage{graphicx}
\usepackage{dcolumn}
\usepackage{bm}
\usepackage{float}
\usepackage{braket}
\usepackage{color}
\begin{document}

\title{Helium-bearing superconductor at high pressure}

\author{Jingyu Hou}
\affiliation{Key Laboratory of Weak-Light Nonlinear Photonics and School of Physics, Nankai University, Tianjin 300071, China}
\affiliation{State Key Laboratory of Metastable Materials Science and Technology, Key Laboratory for Microstructural Material Physics of Hebei Province, School of Science, and Center for High Pressure Science, Yanshan University, Qinhuangdao 066004, China}

\author{Xiao Dong}
\affiliation{Key Laboratory of Weak-Light Nonlinear Photonics and School of Physics, Nankai University, Tianjin 300071, China}

\author{Artem R. Oganov}
\affiliation{Skolkovo Institute of Science and Technology, Bolshoy Boulevard 30, bld. 1, Moscow 121205, Russia}

\author{Xiao-Ji Weng}
\affiliation{Key Laboratory of Weak-Light Nonlinear Photonics and School of Physics, Nankai University, Tianjin 300071, China}
\affiliation{State Key Laboratory of Metastable Materials Science and Technology, Key Laboratory for Microstructural Material Physics of Hebei Province, School of Science, and Center for High Pressure Science, Yanshan University, Qinhuangdao 066004, China}

\author{Chun-Mei Hao}
\affiliation{State Key Laboratory of Metastable Materials Science and Technology, Key Laboratory for Microstructural Material Physics of Hebei Province, School of Science, and Center for High Pressure Science, Yanshan University, Qinhuangdao 066004, China}

\author{Guochun Yang}
\affiliation{State Key Laboratory of Metastable Materials Science and Technology, Key Laboratory for Microstructural Material Physics of Hebei Province, School of Science, and Center for High Pressure Science, Yanshan University, Qinhuangdao 066004, China}

\author{Hui-Tian Wang}
\affiliation{National Laboratory of Solid State Microstructures, School of Physics, and Collaborative Innovation Center of Advanced Microstructures, Nanjing University, Nanjing 210093, China}

\author{Xiang-Feng Zhou}
\email{xfzhou@ysu.edu.cn}
\email{zxf888@163.com}
\affiliation{Key Laboratory of Weak-Light Nonlinear Photonics and School of Physics, Nankai University, Tianjin 300071, China}
\affiliation{State Key Laboratory of Metastable Materials Science and Technology, Key Laboratory for Microstructural Material Physics of Hebei Province, School of Science, and Center for High Pressure Science, Yanshan University, Qinhuangdao 066004, China}

\author{Yongjun Tian}
\affiliation{State Key Laboratory of Metastable Materials Science and Technology, Key Laboratory for Microstructural Material Physics of Hebei Province, School of Science, and Center for High Pressure Science, Yanshan University, Qinhuangdao 066004, China}

\begin{abstract}
Helium (He) is the most inert noble gas at ambient conditions. It adopts a hexagonal close packed structure ($P6_{3}/mmc$) and remains in the insulating phase up to 32 TPa. In contrast, lithium (Li) is one of the most reactive metals at zero pressure, while its cubic high-pressure phase ($Fd\overline{3}m$) is a weak metallic electride above 475 GPa. Strikingly, a stable compound of Li$_{5}$He$_{2}$ ($R\overline{3}m$) was formed by mixing $Fd\overline{3}m$ Li with $P6_{3}/mmc$ He above 700 GPa. The presence of helium promotes the lattice transformation from $Fd\overline{3}m$ Li to $Pm\overline{3}m$ Li, and tuns the three-dimensional distributed interstitial electrons into the mixture of zero- and two-dimensional anionic electrons. This significantly increases the degree of metallization at the Fermi level, consequently, the coupling of conductive anionic electrons with the Li-dominated vibrations is the key factor to the formation of superconducting electride Li$_{5}$He$_{2}$ with a transition temperature up to 26 K, dynamically stable to pressures down to 210 GPa.
\end{abstract}



\maketitle
The element lithium had attracted much attention because of its complex phase diagram and fascinating properties. Li has a body-centered-cubic ($bcc$) structure at ambient conditions and is viewed as a simple metal due to the formation of half-filled nearly-free-electron band and a nearly spherical Fermi surface. \cite{R01,R02,R03,R04} Below the pressure of 100 GPa, lithium undergoes a series of symmetry-breaking phase transitions with the sequence $bcc$ $\rightarrow$ $fcc$ ($\sim$7 GPa) $\rightarrow$ $I\overline{4}3d$ ($\sim$42 GPa) $\rightarrow$ $Aba2$ ($\sim$70 GPa) $\rightarrow$ $Pbca$ ($\sim$80 GPa). \cite{R02,R05} Within this pressure range, the calculated band structures show an unusual electronic transition of metal $\rightarrow$ insulator $\rightarrow$ Dirac semimetal. \cite{R05,R06} As pressure increases, the interatomic distances of materials generally decrease. The valence and conduction bands are thus expected to broaden, leading to pressure-induced metallization. \cite{R07} The abnormal phase transitions for various Li allotropes are mainly due to the distribution of interstitial electrons. Up to $\sim$80 GPa, superconductivity in Li had also been observed while the calculations suggested the semimetallic or even insulating behavior, \cite{R01,R02,R08,R09,R10} and therefore the mechanism for the emergence of superconductivity in Li is not yet fully resolved.

As lithium's neighbor in the periodic table, owing to the closed-shell electronic structure, helium is the most inert noble gas that generally does not interact with other materials at ambient conditions. Moreover, helium remains in the insulating phase up to 32 TPa because of the highest ionization potential ($\sim$25 eV) and zero electron affinity. \cite{R11} However, the crystal structures and electronic properties of materials can be significantly tuned by pressure, leading to the formation of various new materials and complex physical behaviors. For instance, small helium atoms occupy voids in structures and thus increase their packing density, which is helpful to the formation of compounds under pressure, e.g. solid van der Waals materials \cite{R12,R13,R14,R15,R16,R17,R18} or novel ionic compounds. \cite{R19,R20,R21,R22} Recently, a compound of helium and sodium, Na$_{2}$He with a fluorite-type structure, was successfully synthesized at pressure above 113 GPa and discovered to be an insulator with electron pairs localized in interstices. \cite{R07} Since Li is a member of alkali metals, an interesting question arises, that is, whether there are stable lithium-helium compounds under pressure and what are their distinguishing properties.

The variable-composition evolutionary algorithm \textsc{uspex} \cite{R23,R24} was utilized to predict thermodynamically stable compounds in Li-He system. At each pressure, we performed structure searches with an unbiased sampling of the entire range of compositions, varying the stoichiometries and their structures simultaneously. To make the prediction more reliable, two independent searches at every single pressure were performed with the number of atoms per primitive cell ranging from 8 to 24 and from 18 to 40, respectively. For each structure search, the first generation was produced randomly and the fittest 60\% of the population were given the probabilities to be the parents of structures in the next generation $-$ 20\% by heredity, 20\% by lattice mutation, 10\% by transmutation, and 50\% were newly added random structures. \cite{R18} The initial population consisted of 60 structures, all other generations combined add up to $\sim$3000 structures. The total number for variable-composition structure search is $\sim$24,000 at pressures of 250, 500, 800, and 1000 GPa, respectively. Only one stable compound, Li$_{5}$He$_{2}$, was found in these searches. \cite{R07,R25} Structure relaxation and electronic property calculations were carried out within the generalized gradient approximation (GGA) of Perdew-Burke-Ernzerhof (PBE) functional \cite{R26} implemented in the \textsc{vasp} code. \cite{R27} We used the projector-augmented wave method, \cite{R28} with 1$s^{2}$2$s^{1}$ and 1$s^{2}$ electrons of Li and He, respectively, treated as valence. A plane-wave cutoff energy of 600 eV and uniform $\Gamma$-centered $k$ meshes with a resolution of $2 \pi \times 0.06$~\AA$^{-1}$  were used for structure searching, which were further increased to 900 eV and $2 \pi \times 0.025$~\AA$^{-1}$ to guarantee the total energy converges to better than 1 meV/atom. Phonons and electron-phonon coupling (EPC) coefficients were computed on the $7\times 7\times7$ $q$-point meshes via the \textsc{quantum espresso} package. \cite{R29} The Kresse-Joubert projector-augmented waves pseudopotentials \cite{R30} were adopted with the cutoff energy of 110 Ry. The $21\times 21\times 21$ $k$-point meshes in combination with a Methfessel-Paxton smearing \cite{R31} of 0.01 Ry are used to calculate the self-consistent electron densities.

\begin{figure}[t]
\begin{center}
\includegraphics[width=8.5cm]{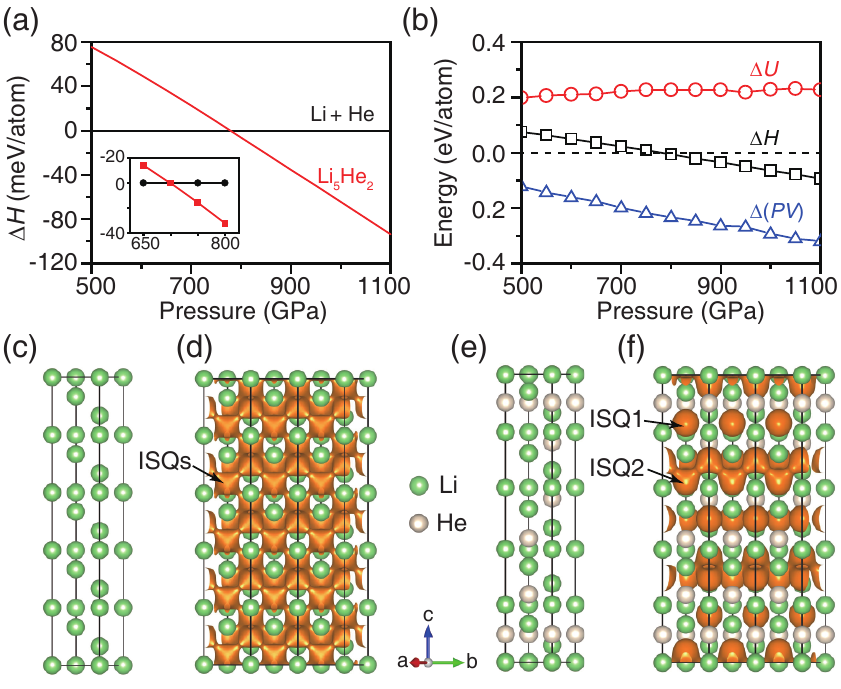}
\caption{\label{f1}
(a) Enthalpy of formation $\Delta$$H$ for $R\overline{3}m$ Li$_{5}$He$_{2}$ as a function of pressure. The inset shows $\Delta$$H$ including the zero-point energy. (b) The evolution of $\Delta$$U$ and $\Delta$($PV$) terms\affiliation{Key Laboratory of Weak-Light Nonlinear Photonics and School of Physics, Nankai University, Tianjin 300071, China} for Li$_{5}$He$_{2}$ with respect to pure elements. (c) and (d) Crystal structure and deformation charge density for sublattice of Li in Li$_{5}$He$_{2}$ ($Pm\overline{3}m$ Li at 800 GPa). The deformation charge density was obtained as the crystal electron density minus superposition of isolated atomic densities. (e) and (f) Crystal structure and the corresponding difference charge density of Li$_{5}$He$_{2}$ at 800 GPa. The interstitial electrons in (d) and (f) were labeled as interstitial quasiatoms (ISQs).}
\end{center}
\end{figure}

In principle, a stable material under pressure has a negative enthalpy of formation with respect to either elements or any other possible compounds. For the structure searches in Li-He system above 500 GPa, the most stable reactants of $Fd\overline{3}m$ Li and $P6_{3}/mmc$ He are used as pure elements, and thus the enthalpy of formation is defined as $\Delta$$H$ = $H$(Li$_{x}$He$_{1-x}$)$-$$x$$H$(Li)$-$(1$-$$x$)$H$(He). The enthalpy $H$ is calculated according to $H$ = $U$ $+$ $PV$, where $U$, $P$, and $V$ represent the internal energy, pressure, and volume, respectively. The results show Li$_{5}$He$_{2}$ is stable above $\sim$778 GPa [Fig.~S1 and Fig.~\hyperref[f1]{1(a)}]. If the calculations include the contribution of zero-point energy, the stabilization pressure decreases to $\sim$700 GPa [Fig.~\hyperref[f1]{1(a)}]. The crystal structure of Li$_{5}$He$_{2}$ belongs to the trigonal crystal system with the space group $R\overline{3}m$. Its hexagonal form is shown in Fig.~\hyperref[f1]{1(e)} with the lattice parameters of $a$ = $b$ = 2.01 {\AA} and $c$ = 12.48 {\AA} at 800 GPa. The He atoms occupies the crystallographic $6c$ sites at (0.000, 0.000, 0.097), and the Li atoms occupies $6c$ sites with coordinates (0.000, 0.000, 0.607), (0.000, 0.000, 0.800), and $3a$ sites at (0.000, 0.000, 0.000). The sublattice of Li has a simple cubic structure ($Pm\overline{3}m$) with the transformation matrix $\begin{pmatrix} 1&\overline{1}&0 \\ 0&1&\overline{1} \\ 5&5&5 \\ \end{pmatrix}$, which matches well with the rhombohedral-centered substructure of He [Fig.~\hyperref[f1]{1(c)}]. To reveal the origin of thermodynamical stability in Li$_{5}$He$_{2}$, the evolution of $\Delta$$U$ and $\Delta$($PV$) terms as a function of pressure was plotted in Fig.~\hyperref[f1]{1(b)}. It shows that the value of $\Delta$$U$ increases slightly (approximately remained constant) while $\Delta$($PV$) dramatically decreases as the pressure increases. The presence of helium increases the packing density, reducing the enthalpy of formation and results in the formation of Li$_{5}$He$_{2}$ above $\sim$700 GPa. The next step is to understand which role helium plays in the electronic structure.

Electrides represent a class of exotic compounds where valence electrons reside at interstices of a host structure and behave as anionic quasiatoms, which significantly determines their properties. \cite{R32,R33} According to the distribution and dimensionality of anionic electrons, electrides can be classified into zero-dimensional (0D), one-dimensional (1D), two-dimensional (2D), and three-dimensional (3D) ones. \cite{R34} At high pressure, $Fd\overline{3}m$ Li consists of Li ions arranged in the diamond structure. In this structure, the interstitial electrons and the Li ions form interpenetrating diamond lattices, taking the Li and electron sites together, they form a 3D electride state with Zintl structure (NaTl-type) above 475 GPa. \cite{R01} As mentioned above, $R\overline{3}m$ Li$_{5}$He$_{2}$ can be formed by mixing $Fd\overline{3}m$ Li with $P6_{3}/mmc$ He above 700 GPa, and therefore it is intriguing to study its electronic properties. Interestingly, the sublattice of Li in Li$_{5}$He$_{2}$ ($Pm\overline{3}m$ Li) is also a 3D electride, in which the interstitial electrons occupy the body center of the lattice and interconnect with each other [Fig.~\hyperref[f1]{1(d)} and Fig.~S2(a)]. However, the sublattice of He in Li$_{5}$He$_{2}$ alters the number and distribution of interstitial electrons in $Pm\overline{3}m$ Li, leading to the formation of two inequivalent interstitial quasiatoms (ISQs) which occupy the $3b$ sites (termed ISQ1) at (0.000, 0.000, 0.500) and $6c$ sites (termed ISQ2) at (0.000, 0.000, 0.300). The nearest distance of Li-ISQ1-Li and Li-ISQ2-Li are 2.67 {\AA} and 2.41 {\AA} at 800 GPa whereas the average bond length of Li-Li is 1.43 {\AA}. The estimated radii of ISQ1 and ISQ2 are $\sim$0.62 and $\sim$0.50 {\AA}, respectively. Therefore, as shown in Fig.~\hyperref[f1]{1(f)} and Fig.~S2(b), it looks that ISQ1 is larger in size than ISQ2. Meanwhile, both the deformation charge density and electron localization function (ELF) confirmed that ISQ1 possess cage states because they are localized in the isolated cavities while ISQ2 are interconnected to form a puckered 2D layer [Fig.~\hyperref[f1]{1(f)} and Fig.~S2(b)]. Consequently, Li$_{5}$He$_{2}$ is a peculiar ionic compound featured by the coexistence of 0D and 2D anionic electrons. In addition, Bader analysis of the total electron density was performed to investigate the charge transfer of Li$_{5}$He$_{2}$ at pressures of 700, 800, and 900 GPa. \cite{R25,R35} The results show a large amount of charge is transferred from Li to ISQs while a small amount of charge to He [Table S\hyperref[t1]{I}]. As a whole, the electride of Li$_{5}$He$_{2}$ at 800 GPa, could be designated as [Li$_{5}$He$_{2}$]$^{2.35+}$:2.35$e^{-}$.

\begin{figure}[t]
\begin{center}
\includegraphics[width=8.5cm]{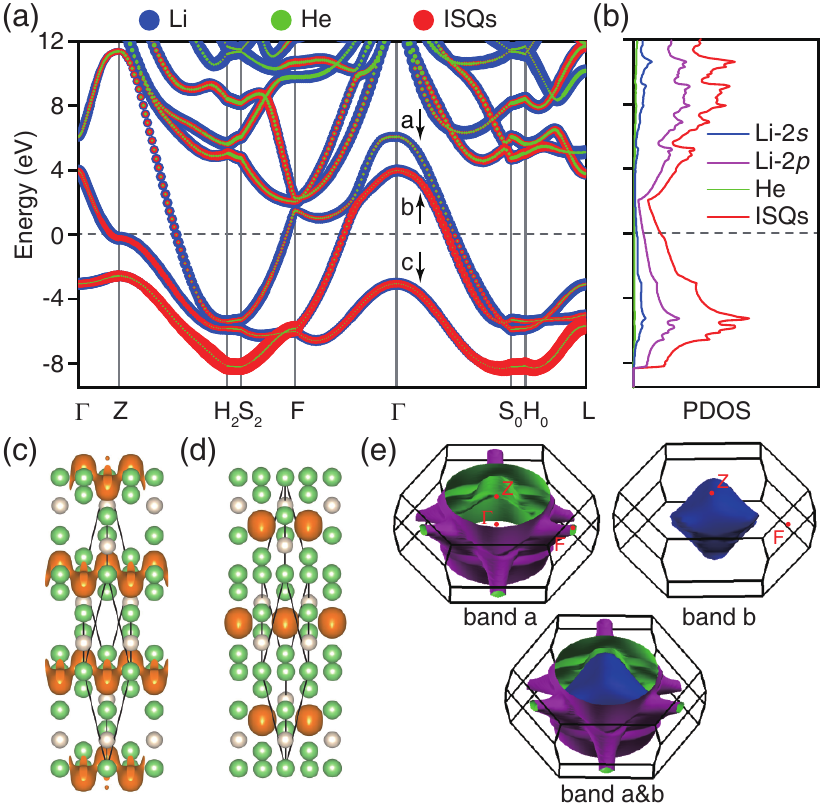}
\caption{\label{f2}
 Electronic properties of $R\overline{3}m$ Li$_{5}$He$_{2}$ at 800 GPa. (a) The orbital-resolved band structures. (b) Projected density of states (PDOS). PDOS for ISQs was calculated by projecting interstitial electrons onto virtual orbitals. The estimated radii of ISQ1 and ISQ2 are $\sim$0.6 {\AA} and 0.5 {\AA}, respectively. (c) The band-decomposed charge density of Li$_{5}$He$_{2}$ with the energy range from $-$0.25 to 0.25 eV and (d) from $-$8.5 to $-$7.0 eV. (e) Fermi surface of Li$_{5}$He$_{2}$.}
\end{center}
\end{figure}

The orbital-resolved band structure shows that the $R\overline{3}m$ Li$_{5}$He$_{2}$ is metallic because its two bands cross the Fermi level $E$${\rm_{F}}$, marked as band-a and band-b, respectively [Fig.~\hyperref[f2]{2(a)}]. The band-a is dominantly derived from the Li $p$ orbitals, arising from the $s$ $\rightarrow$ $p$ electronic transition of Li under pressure (see the projected density of states) [Fig.~\hyperref[f2]{2(b)}]. By comparison, the band-b is mostly derived from the states of ISQs. The normalized electronic DOS at $E$${\rm_{F}}$ is $\sim$0.021 states/eV for Li$_{5}$He$_{2}$ at 800 GPa, whereas $\sim$0.013 states/eV for $Fd\overline{3}m$ Li, suggesting the degree of metallization of Li$_{5}$He$_{2}$ is increased by helium-tuned ISQs. In particular, the band-decomposed charge density with the energy range from $-$0.25 to 0.25 eV shows that ISQ2 makes the greatest contribution to the conductivity of Li$_{5}$He$_{2}$. In contrast, the band-decomposed charge density with energy range from $-$8.5 to $-$7.0 eV shows that ISQ1 is mainly responsible for the less dispersive band-c. In addition, there are sharp PDOS peaks and strong hybrids within the energy range from $-$6.5 to $-$3.0 eV, implying the major electrostatic interaction between Li 2$p$ states and ISQ1 cage states. All these are related to the cage states of ISQ1, indicating that the localized states of ISQ1 make no contribution to the conductivity of Li$_{5}$He$_{2}$. The corresponding Fermi surface is plotted in Fig.~\hyperref[f2]{2(e)}. The band-a makes the toroidal and multiterminal tubelike hole pockets (i.e., around the high-symmetry point of F) while band-b makes the irregular cage-like electron pockets (e.g., around the high-symmetry point of Z). A good Fermi surface nesting appears in Li$_{5}$He$_{2}$ along $\Gamma$ $\rightarrow$ F line with highly dispersive bands along this direction. Therefore, the band structure reveals a good metallicity with large dispersion bands crossing $E$${\rm_{F}}$ and a relatively flat band in the vicinity of $E$${\rm_{F}}$ close to the Z point. The coexistence of steep and flat bands near $E$${\rm_{F}}$ implies a favorable condition for enhancing the formation of Cooper pairs by providing a vanishing Fermi velocity to part of the conduction electrons, \cite{R36} which is essential to the superconductivity of Li$_{5}$He$_{2}$ under pressures. \cite{R37,R38}

\begin{figure}[t]
\begin{center}
\includegraphics[width=8.5cm]{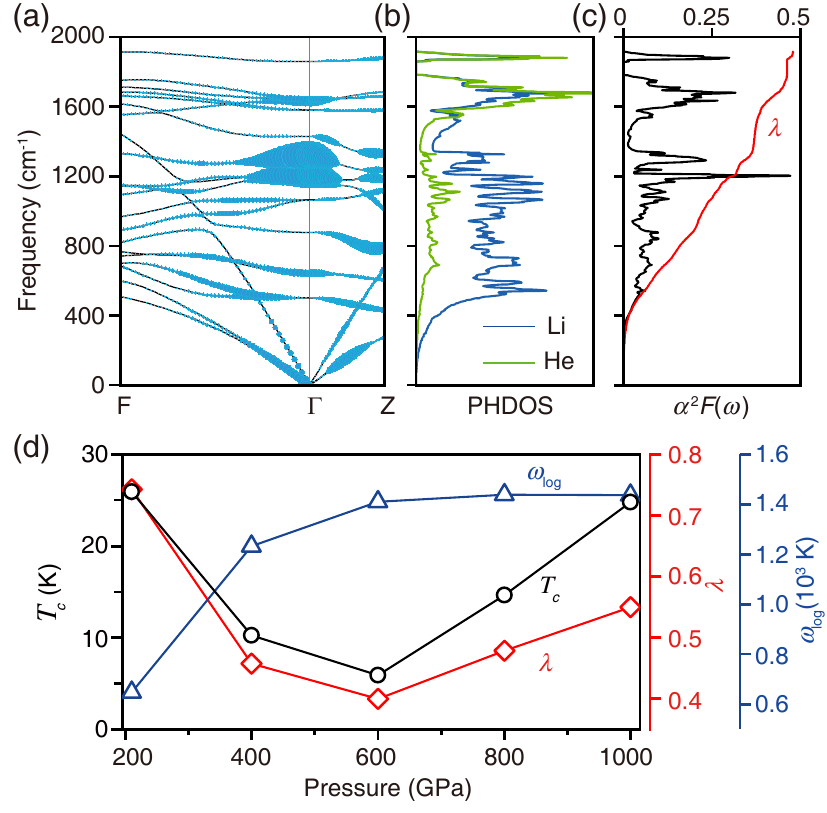}
\caption{\label{f3}
Superconducting properties of $R\overline{3}m$ Li$_{5}$He$_{2}$ at high pressures. (a)-(c) Phonon dispersion curve, projected phonon DOS, Eliashberg spectral function $\alpha^{2}F(\omega)$, and EPC parameter $\lambda$ of Li$_{5}$He$_{2}$ at 800 GPa. The area of blue solid circles is proportional to the partial EPC $\lambda_{q,v}$. (d) The EPC $\lambda$, the logarithmic average phonon frequency $\omega$${\rm_{log}}$, and the critical temperature $T_{c}$ were plotted as a function of pressure. }
\end{center}
\end{figure}

The phonon dispersion curves in conjunction with EPC at different pressures were calculated to investigate the dynamical stability and superconductivity of Li$_{5}$He$_{2}$. The superconducting transition temperature ($T_{c}$) was estimated by the Allen-Dynes-modified McMillan equation \cite{R39,R40} $$T_{c}=\frac{\omega{\rm_{log}}}{1.2}exp\left[-\frac{1.04(1+\lambda)}{\lambda-\mu^{*}(1+0.62\lambda)}\right],$$ where $\lambda$ is the EPC strength, $\omega$${\rm_{log}}$ is the logarithmic average phonon frequency, and $\mu^{*}$ is the Coulomb pseudopotential parameter. The parameters $\lambda$ and $\omega$${\rm_{log}}$ are defined as $$\lambda=2\int_0^{\infty}\frac{\alpha^{2}F(\omega)}{\omega}\,d\omega$$ and $$\omega{\rm_{log}}=exp\left[\frac{2}{\lambda}\int_0^{\infty}\frac{d\omega}{\omega}\alpha^{2}F(\omega)ln\omega\right],$$ respectively. Here a typical value of $\mu^{*}$= 0.1 was used for the calculation of $T_{c}$. The absence of imaginary phonon frequencies in the whole Brillouin zone indicates that it is dynamically stable in the pressure range from 210 to at least 1000 GPa (Fig.~\hyperref[f3]{3} and Fig.~S3). For instance, the calculated $\lambda$ and $\omega$${\rm_{log}}$ at 800 GPa are 0.48 and 1438 K, thus the predicted $T_{c}$ is equal to $\sim$15 K. Note that the values of $T_{c}$ are decreased as $\mu^{*}$ is increased (Fig.~S4) whereas the $U$-shaped behavior of $T_{c}$ is reserved. \cite{R41} These calculations prove that $R\overline{3}m$ Li$_{5}$He$_{2}$ is the phonon-mediated superconducting electride. By contrast, the superconducting properties of reference phase $Fd\overline{3}m$ Li were also calculated by using the same method at pressures of 600, 800, and 1000 GPa. As shown in Table S\hyperref[t1]{II}, both the parameters of $\lambda$ and $\omega$${\rm_{log}}$ of $Fd\overline{3}m$ Li are lower than those of Li$_{5}$He$_{2}$ at the corresponding pressures. \cite{R25} The predicted $T_{c}$ of $Fd\overline{3}m$ Li is less than 1 K from 600 to 1000 GPa, indicating it is a very weak superconductor at high pressures. Since helium is the most inert noble gas at ambient conditions, it is very interesting to explore helium's contribution to the superconductivity of Li$_{5}$He$_{2}$ under pressure. As shown in Figs.~\hyperref[f3]{3(a)-3(c)}, the phonon dispersion curves associated with the partial EPC $\lambda_{q,v}$ suggest that almost all the phonon modes contribute to the EPC strength $\lambda$. However, there is also a distinct character in the projected phonon DOS (PHDOS), that is, the Li atoms dominate the vibrations below 1550 cm$^{-1}$ while Li and He contribute to the coupling vibration modes above 1550 cm$^{-1}$. In general, the PHDOS can be divided into two parts: One is the Li-dominated vibrations and the other is the coupled vibrations. According to this definition, these two parts contribute 77$\%$ and 23$\%$ to the EPC strength $\lambda$, respectively. As pressure is varied, the values of $T_{c}$ show a non-monotonic dependence with $T_{c}$ decreasing from $\sim$26 K at 210 GPa to $\sim$6 K (600 GPa), and then increasing to $\sim$25 K at 1000 GPa. To clarify such unusual superconducting behavior, the pressure-dependent superconducting properties of Li$_{5}$He$_{2}$ were plotted in Fig.~\hyperref[f3]{3(d)}. One can see that $\omega$${\rm_{log}}$ increases rapidly up to 600 GPa and then remains almost a constant, while $\lambda$ shows a similar trend to $T_{c}$. Based on the PHDOS and Eliashberg phonon spectral function $\alpha^{2}F(\omega)$ at different pressures (Fig.~S3 and Table \hyperref[t1]{I}), the contributions to EPC $\lambda$ from the Li-dominated vibrations are $\sim$67$\%$ both at 210 GPa and 400 GPa, 68$\%$ at 600 GPa, 77$\%$ at 800 GPa, and 85$\%$ at 1000 GPa, respectively. \cite{R25} Obviously, the Li-dominated vibrations play a decisive role in the superconductivity of Li$_{5}$He$_{2}$ at high pressures.

\begin{table}[t]
\caption{\label{t1}
Total EPC $\lambda$, partial $\lambda{\rm_{i}}$ contributed by Li-dominated vibrations, and partial $\lambda{\rm_{ii}}$ contributed by the the coupling vibrations between Li and He at different pressures, note that $\lambda$=$\lambda{\rm_{i}}$+$\lambda{\rm_{ii}}$.}
\renewcommand\arraystretch{1.3}{
\setlength{\tabcolsep}{0.6cm}{
\resizebox{\linewidth}{!}{
\begin{tabular}{cccc}
\hline\hline
Pressure (GPa) & $\lambda{\rm_{i}}$ & $\lambda{\rm_{ii}}$ & $\lambda$ \\
\hline
210 & 0.50 & 0.25 &0.75 \\
400 & 0.31 & 0.15 &0.46 \\
600 & 0.27 & 0.13 &0.40 \\
800 & 0.37 & 0.11 &0.48 \\
1000& 0.47 & 0.08 &0.55 \\
\hline\hline
\end{tabular}}}}
\end{table}

In conclusion, the high-pressure phases of Li-He system were systematically investigated by the \textit{ab initio} evolutionary searches. We predicted that there is only one stable compound, Li$_{5}$He$_{2}$, which is thermodynamically stable above 700 GPa. The first-principle calculations reveal that Li$_{5}$He$_{2}$ is an exotic electride with the coexistence of 0D and 2D anionic electrons. Further EPC calculations identified Li$_{5}$He$_{2}$ as a phonon-mediated superconducting electride with $T_{c}$ up to 26 K. This prediction establishes the first helium-bearing superconductor at high pressure, which greatly enriches the systems and types of superconducting materials.

This work was supported by the National Natural Science Foundation of China (Grants No. 52025026, 11874224, and 52090020), and the Young Elite Scientists Sponsorship Program by Tianjin (Grant No. TJSQNTJ-2018-18). A.R.O. acknowledges funding from the Russian Science Foundation (Grant No. 19-72-30043).




\begin{references}
\bibitem{R01} C. J. Pickard and R. J. Needs, Phys. Rev. Lett. \textbf{102}, 146401 (2009).

\bibitem{R02} J. Lv, Y. Wang, L. Zhu, and Y. Ma, Phys. Rev. Lett. \textbf{106}, 015503 (2011).

\bibitem{R03} C. J. Pickard and R. J. Needs, Phys. Rev. Lett. \textbf{107}, 087201 (2011).

\bibitem{R04} A. M. J. Schaeffer, W. B. Talmadge, S. R. Temple, and S. Deemyad, Phys. Rev. Lett. \textbf{109}, 185702 (2012).

\bibitem{R05} S. A. Mack, S. M. Griffin, and J. B. Neaton, Proc. Natl. Acad. Sci. USA \textbf{116}, 9197 (2019).

\bibitem{R06} S. F. Elatresh, Z. Zhou, N. W. Ashcroft, S. A. Bonev, J. Feng, and R. Hoffmann, Phys. Rev. Mater. \textbf{3}, 044203 (2019).

\bibitem{R07} X. Dong, A. R. Oganov, A. F. Goncharov, E. Stavrou, S. Lobanov, G. Saleh, G. R. Qian, Q. Zhu, C. Gatti, V. L. Deringer, R. Dronskowski, X. F. Zhou, V. B. Prakapenka, Z. Kon\^opkov\'a, I. A. Popov, A. I. Boldyrev, and H. T. Wang, Nat. Chem. \textbf{9}, 440 (2017).

\bibitem{R08} V. V. Struzhkin, M. I. Eremets, W. Gan, H. k. Mao, and R. J. Hemley, Science \textbf{298}, 1213 (2002).

\bibitem{R09} K. Shimizu, H. Ishikawa, D. Takao, T. Yagi, and K. Amaya, Nature (London) \textbf{419}, 597 (2002).

\bibitem{R10} J. B. Neaton and N. W. Ashcroft, Nature (London) \textbf{400}, 141 (1999).

\bibitem{R11} B. Monserrat, N. D. Drummond, C. J. Pickard, and R. J. Needs, Phys. Rev. Lett. \textbf{112}, 055504 (2014).

\bibitem{R12} W. L. Vos, L. W. Finger, R. J. Hemley, J. Z. Hu, H. K. Mao, and J. A. Schouten, Nature (London) \textbf{358}, 46 (1992).

\bibitem{R13} Y. Li, X. Feng, H. Liu, J. Hao, S. A. T. Redfern, W. Lei, D. Liu, and Y. Ma, Nat. Commun. \textbf{9}, 722 (2018).

\bibitem{R14}  D. Li, Y. Liu, F. Tian, S. Wei, Z. Liu, D. Duan, B. Liu, and T. Cui, Solid State Commun. \textbf{283}, 9 (2018).

\bibitem{R15} C. Liu, H. Gao, Y. Wang, R. J. Needs, C. J. Pickard, J. Sun, H. T. Wang, and D. Xing, Nat. Phys. \textbf{15}, 1065 (2019).

\bibitem{R16} C. Liu, H. Gao, A. Hermann, Y. Wang, M. Miao, C. J. Pickard, R. J. Needs, H. T. Wang, D. Xing, and J. Sun, Phys. Rev. X \textbf{10}, 021007 (2020).

\bibitem{R17} J. Shi, W. Cui, J. Hao, M. Xu, X. Wang, and Y. Li, Nat. Commun. \textbf{11}, 3164 (2020).

\bibitem{R18} J. Hou, X. J. Weng, A. R. Oganov, X. Shao, G. Gao, X. Dong, H. T. Wang, Y. Tian, and X. F. Zhou, Phy. Rev. B \textbf{103}, L060102 (2021).

\bibitem{R19} B. Monserrat, M. Martinez-Canales, R. J. Needs, and C. J. Pickard, Phys. Rev. Lett. \textbf{121}, 015301 (2018).

\bibitem{R20} Z. Liu, J. Botana, A. Hermann, S. Valdez, E. Zurek, D. Yan, H. Q. Lin, and M.-S. Miao, Nat. Commun. \textbf{9}, 951 (2018).

\bibitem{R21} J. Zhang, J. Lv, H. Li, X. Feng, C. Lu, S. A. T. Redfern, H. Liu, C. Chen, and Y. Ma, Phys. Rev. Lett. \textbf{121}, 255703 (2018).

\bibitem{R22} H. Gao, J. Sun, C. J. Pickard, and R. J. Needs, Phys. Rev. Mater. \textbf{3}, 015002 (2019).

\bibitem{R23} A. R. Oganov and C. W. Glass, J. Chem. Phys. \textbf{124}, 244704 (2006).

\bibitem{R24} O. Lyakhov, A. R. Oganov, H. T. Stokes, and Q. Zhu, Comput. Phys. Commun. \textbf{184}, 1172 (2013).

\bibitem{R25} See Supplemental Material at http://link.aps.org/supplemental/ for an additional calculations and electronic properties.

\bibitem{R26} J. P. Perdew, K. Burke, and M. Ernzerhof, Phys. Rev. Lett. \textbf{77}, 3865 (1996).

\bibitem{R27} G. Kresse and J. Furthm\"{u}ller, Phys. Rev. B \textbf{54}, 11169 (1996).

\bibitem{R28} P. E. Bl\"ochl, Phys. Rev. B \textbf{50}, 17953 (1994).

\bibitem{R29} P. Giannozzi, S. Baroni, N. Bonini, M. Calandra, R. Car, C. Cavazzoni, D. Ceresoli, G. L. Chiarotti, M. Cococcioni, and I. Dabo, J. Phys. Condens. Matter \textbf{21}, 395502 (2009).

\bibitem{R30} G. Kresse and D. Joubert, Phys. Rev. B \textbf{59}, 1758(1999).

\bibitem{R31} M. Methfessel and A. T. Paxton, Phys. Rev. B \textbf{40}, 3616 (1989).

\bibitem{R32} Z. Zhao, S. Zhang, T. Yu, H. Xu, A. Bergara, and G. Yang, Phys. Rev. Lett. \textbf{122}, 097002 (2019).

\bibitem{R33} Z. Liu, Q. Zhuang, F. Tian, D. Duan, H. Song, Z. Zhang, F. Li, H. Li, D. Li, and T. Cui, Phys. Rev. Lett. \textbf{127}, 157002 (2021).

\bibitem{R34} Y. Tsuji, P. L. V. K. Dasari, S. F. Elatresh, R. Hoffmann, and N. W. Ashcroft, J. Am. Chem. Soc. \textbf{138}, 14108(2016).

\bibitem{R35} W. Tang, E. Sanville, and G. Henkelman, J. Phys. Condens. Matter \textbf{21}, 084204 (2009).

\bibitem{R36} A. Simon, Angew. Chem. Int. Ed. Engl. \textbf{36}, 1788 (1997).

\bibitem{R37} J. S. Tse, Y. Yao, and K. Tanaka, Phys. Rev. Lett. \textbf{98}, 117004 (2007).

\bibitem{R38} X. F. Zhou, A. R. Oganov, X. Dong, L. Zhang, Y. Tian, and H. T. Wang, Phys. Rev. B \textbf{84}, 054543 (2011).

\bibitem{R39} P. B. Allen and R. C. Dynes, Phys. Rev. B \textbf{12}, 905 (1975).

\bibitem{R40} P. B. Allen, Phys. Rev. Lett. \textbf{59}, 1460 (1987).

\bibitem{R41} G. Profeta, C. Franchini, N. N. Lathiotakis, A. Floris, A. Sanna, M. A. L. Marques, M. L\"uders, S. Massidda, E. K. U. Gross, and A. Continenza, Phys. Rev. Lett. \textbf{96}, 047003 (2006).

\end{references}
\end{document}